\documentclass[aps,pre,twocolumn,showpacs,superscriptaddress]{revtex4}

\usepackage{graphicx}
\usepackage{dcolumn}
\usepackage{bm}

\def\be{\begin{equation}}
\def\ee{\end{equation}}
\def\bea{\begin{eqnarray}}
\def\eea{\end{eqnarray}}
\def\bi{\begin{itemize}}
\def\ei{\end{itemize}}

\begin{document}
\preprint{\today}

\title{Local virial relation for self-gravitating system}

\author{Osamu Iguchi \footnote{osamu@phys.ocha.ac.jp}}
  \affiliation{Department of Physics, Ochanomizu University,
	       2-1-1 Ohtuka, Bunkyo-ku, Tokyo 112-8610, Japan}

\author{Yasuhide Sota \footnote{sota@cosmos.phys.ocha.ac.jp}}
  \affiliation{Department of Physics, Ochanomizu University,
	       2-1-1 Ohtuka, Bunkyo, Tokyo 112-8610, Japan}
  \affiliation{Advanced Research Institute for Science and Engineering, 
               Waseda University, Ohkubo, Shinjuku-ku, Tokyo 169-8555, Japan}

\author{Akika Nakamichi \footnote{akika@astron.pref.gunma.jp}}
  \affiliation{Gunma Astronomical Observatory,
               6860-86, Nakayama, Takayama, Agatsuma, Gunma 377-0702, Japan}

\author{Masahiro Morikawa \footnote{hiro@phys.ocha.ac.jp}}
  \affiliation{Department of Physics, Ochanomizu University,
               2-1-1 Ohtuka, Bunkyo-ku, Tokyo 112-8610, Japan}

\begin{abstract}
We demonstrate that the quasi-equilibrium state 
in self-gravitating $N$-body system after cold collapse 
are uniquely characterized by the local virial relation 
using numerical simulations.
Conversely assuming the constant local virial ratio and Jeans equation for 
spherically steady state system, 
we investigate the full solution space of the problem 
under the constant anisotropy parameter and 
obtain some relevant solutions.
Especially, 
the local virial relation always provides 
a solution which has a power law density profile 
in both the asymptotic regions $r\rightarrow 0$ and $\infty$.
This type of solutions observed commonly in many numerical simulations.
Only the anisotropic velocity dispersion controls 
this asymptotic behavior of density profile.

\end{abstract}


\maketitle


\section{\label{sec:intro}Introduction}

Galaxies and clusters of galaxies are typical 
self-gravitating systems (SGS) in the universe.
Extracting the essence of them, 
study on the quasi-equilibrium state of SGS is 
an important clue for understanding 
the basic properties of structures in the universe. 
In the process of the structure formation of SGS, 
a cold collapse and a cluster-pair collision would be 
the most fundamental processes.
Therefore in this paper, 
we would like to focus on the characteristic properties
of the quasi-equilibrium state formed by such basic processes.

For many numerical simulations, 
after a cold collapse, 
system settles down to a quasi-equilibrium state which is well described by 
the Vlasov equation for a spherically symmetric steady state system.
In such a quasi-equilibrium state, 
the spherically averaged density profile is found 
to be $\rho \propto r^{-4}$ at the outer region\cite{Henon64,Albada82} and 
the velocity dispersion is isotropic near the central region and 
becomes radially anisotropic at the outer region\cite{Binney87}.
Recently we found two characteristic properties which appearer 
in the quasi-equilibrium state after a cold collapse\cite{Iguchi04,Sota04}.
One is the linear temperature-mass relation which yields a
characteristic non-Gaussian velocity distribution.
The other is the local virial (LV) relation 
which is robust against various initial condition such as 
the density profile and the virial ratio.

The Plummer model is one of the examples which 
obey the LV relation.
N.W.Evans and J.An discuss the anisotropic distribution function 
where the LV relation is satisfied ( they call it hypervirial ) and 
obtain the analytical solution 
which has a constant anisotropy parameter\cite{Evans05}.
In this paper, 
we would like to explore the relevance of LV relation in 
dynamics of SGS.
Therefore, 
we assume the general LV relation where the virial ratio is constant and 
investigate the role of LV relation in a spherically symmetric steady state system.
Moreover, 
we compare the solutions of Jeans equation under the LV relation 
and the results of numerical simulations.
In numerical simulations, 
we use a leap-frog symplectic integrator on GRAPE-5, 
a special-purpose computer designed to accelerate N-body simulations\cite{GRAPE}.

In section \ref{sec:LV}, 
we first show the validity of LV relation 
with the numerical N-body simulation.
Then, assuming the constant virial ratio, 
we investigate the full solution space of Jeans equation 
under the constant anisotropy parameter in section \ref{sec:Jeans}.
We show the special properties of the LV relation 
and obtain analytically the physically relevant solutions.
In section \ref{sec:comparison},
we compare these solutions with the results of numerical simulation 
and discuss the validity and the consistency of the LV relation.
The last section \ref{sec:con} is devoted to the discussions 
and further developments of the present work.

\section{\label{sec:LV}Local virial relation}

In a steady state, 
the gravitationally bound system
settles down to a virialized state satisfying the condition 
\be
\overline{W}+2\overline{K}=0,
\label{Virial}
\ee
where $\overline{W}$ and $\overline{K}$ are, respectively, 
the time-averaged potential energy and kinetic energy 
of the whole bound system. 
This is a well-known global relation that holds for the entire system.

\begin{figure} 
\begin{center}
\includegraphics[width=8cm]{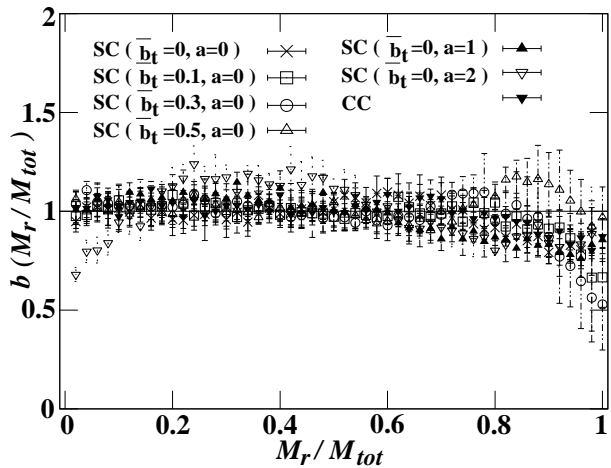}
\end{center}
\caption{\label{fig-cold}
The LV relation for some numerical simulations
obtained by a typical cold collapse simulation;
spherical collapse (SC) and cluster-pair collisions (CC). 
In the case of SC, 
$5000$ particles are distributed with a power law density profile 
($\protect\rho\propto r^{-a}$) within a sphere of radius $R$ 
and the initial virial ratio ($\bar{b}_t$) is set to be small. 
In the case of CC, each cluster has 
the equal number of particles ($2500$) 
and all particles are homogeneously distributed within a sphere of radius $R$ 
and is set to be virialized initially. 
The initial separation of the pair is $6R$ along the $x$ axis. 
In all of the simulations, 
softening length $\protect\epsilon=2^{-8}R$ is introduced to 
reduce the numerical error caused by close encounters.
The LV ratio $b$ is plotted 
as a function of $M_{r}/M_{tot}$, where $M_{tot}$ is 
the total mass of the system.
The virial ratios at $M_r$ are time averaged.}
\end{figure}

Here we define the LV relation at each position $r$ as
\be
2\sigma^{2}(r)=-\Phi (r),
\label{LV-relation}
\ee
where $\sigma$ and $\Phi$ are the velocity dispersion
($\sigma^2=\sigma_r^2+\sigma_\theta^2+\sigma_\phi^2$)  and 
the potential energy, respectively.
Hence, we define the LV ratio 
\be
b(r) := -2\sigma^{2}(r)/\Phi(r),
\label{2k/W}
\ee
in order to measure to what extent the LV relation holds.
Actually, 
this LV relation appearers in the quasi-equilibrium state
obtained by some numerical simulations \cite{Iguchi04,Sota04}.
For a cold collapse and cluster-pair collision 
from a initially smooth density profile,
the LV ratio is shown in Fig.\ref{fig-cold}.
The value $b(r)$ takes almost unity in the inner region 
and slightly reduces in outer region for all of the simulations.
On the other hand, 
for the case of SC($\overline{b_t}=0, a=2$) 
where the initial virial ratio $\overline{b_t}=0$ and 
the initial density profile $\rho\propto r^{-2}$, 
the deviation at the central region is significant.

For a wide class of collapses from various initial conditions 
such as the change of the virial ratio and the density profile,
the LV relation are almost satisfied. 
In the next section,
we investigate the role of the LV relation 
in a steady state of the SGS.

\section{\label{sec:Jeans}Local virial relation and Jeans equation}

To begin with, 
we assume the general LV relation, 
\be
\frac{2\sigma^2(r)}{\phi(r)}=b,
\label{GLV}
\ee
where the relative potential $\phi(r):=-\Phi(r)$ and $b$ is a constant.
When $b=1$, 
LV relation is exactly satisfied.
The Jeans equation for a spherically symmetric steady state system is
\be
\frac{d(\rho\sigma_r^2)}{dr}+\frac{2\beta}{r}\rho\sigma^2_{r}=\rho \frac{d\phi}{dr},
\label{Jeans}
\ee
where the anisotropy parameter $\beta$ is defined as
\be
\beta:=1-\frac{\sigma_\theta^2+\sigma_\phi^2}{2\sigma_r^2}.
\ee 
We study the density profile 
which satisfies the general LV relation (\ref{GLV}) and 
Jeans equation for a spherically symmetric steady state system (\ref{Jeans}).
We consider the simplest case, where $\beta$ is a constant. 
Following W.~Dehnen and D.E.~McLaughlin \cite{Dehnen05}, 
we investigate the full solution space of the problem.

Eliminating $\sigma_r^2$ from Eq.(\ref{Jeans}) by using Eq.(\ref{GLV}), 
the relative potential becomes
\be
\phi=\phi_0 x^{2\beta/\alpha}y^{1/\alpha},
\label{Phi-rho}
\ee
where 
$x:=r/r_0$, $y:=\rho/\rho_0$, 
$\alpha:=(6-4\beta-b)/b$, and $\phi_0:=\phi(r_0)$.
Substituting Eq.(\ref{Phi-rho}) into the Poisson equation, 
the density profile satisfies the following equation:
\be
\gamma^{\prime}-\frac{1}{\alpha}(\gamma-\gamma_a)(\gamma-\gamma_b)=
\alpha\kappa x^{2-2\beta/\alpha}y^{1-1/\alpha},
\label{gamma-eq1}
\ee
where a prime denotes the differentiation with respect to $\ln x$ and 
$\gamma:=-d\ln y/d\ln x$ and $\kappa:=4\pi Gr_0^2\rho_0/\phi_0$, and 
\be
\gamma_a:=2\beta, \quad\quad
\gamma_b:=2\beta+\alpha.
\ee
We will restrict our consideration to the case with $\beta\le 1$ and $0<b<2$.

First, 
the equation (\ref{gamma-eq1}) has 
a scale invariant solution $y=x^{-\gamma_1}$ with 
\be
\gamma_1=\frac{2(\alpha-\beta)}{\alpha-1},
\label{gamma1}
\ee
for $\alpha\neq 1$.
In the case for $\alpha=1$, 
Eq.(\ref{gamma-eq1}) has no scale invariant solution 
except for $(\beta,b)=(1,1)$.
For $(\beta,b)=(1,1)$,  
the r.h.s of Eq.(\ref{gamma-eq1}) becomes always constant ($\kappa$) 
and we can easily integrate the Eq.(\ref{gamma-eq1}) and 
the total mass is infinite ({\it see} Appendix \ref{sec:alpha1}).
Hereafter we consider the case for $\alpha\neq 1$ 
(we discuss the case for $\alpha=1$ in Appendix \ref{sec:alpha1}).
In this case, 
the structure of the phase space ($\gamma\prime, \gamma$) is 
classified into three cases.
\begin{itemize}
\item[(A)]
$\gamma_1<\gamma_a<\gamma_b$ for 
\bea
\frac{3-b}{2} < \beta < 1
\quad(1<b<2),
\nonumber
\eea

\item[(B)]
$\gamma_a<\gamma_1<\gamma_b$ for
\bea
\beta < 1 (0<b\le 1) 
\quad\mbox{or}\quad
\beta < \frac{3-2b}{2-b}
(1<b<2),
\nonumber
\eea

\item[(C)]
$\gamma_a<\gamma_b<\gamma_1$ for
\bea
\frac{3-2b}{2-b}< \beta < \frac{3-b}{2}
\quad(1<b<2).
\nonumber
\eea

\end{itemize}
For each case, 
the asymptotic behavior of $\gamma$ is summarized 
in Table \ref{tab:asympto}.
Note that the solution with the power law behavior 
in both of the asymptotic regions
($r\rightarrow 0,\infty$) exists only in the case B.
Since $\kappa>0$ and $\alpha>0$, 
non-negativity of the r.h.s of Eq.(\ref{gamma-eq1}) leads to 
the following condition:
\be
\gamma^{\prime}\ge\frac{1}{\alpha}(\gamma-\gamma_a)(\gamma-\gamma_b).
\label{gamma-condition}
\ee
From the condition (\ref{gamma-condition}), 
both of the case A and case C have no scale invariant solution ($\gamma_1$).

\begin{table}[tbp]
\caption{The asymptotic behavior of $\gamma$ in Eq.(\ref{gamma-eq1}).
}
\label{tab:asympto}%
\begin{ruledtabular}
\begin{tabular}{lcc}
case
& $r\rightarrow 0$
& $r\rightarrow \infty$\\\hline
A  & $\gamma\rightarrow \gamma_a$, $\gamma_b$, or $-\infty$ 
   & $\gamma\rightarrow \infty$ \\
B  & $\gamma\rightarrow \gamma_a$, $\gamma_1$, or $-\infty$ 
   & $\gamma\rightarrow \gamma_b$, $\gamma_1$, or $\infty$ \\
C  & $\gamma\rightarrow \gamma_a$, $\gamma_b$, or $-\infty$ 
   & $\gamma\rightarrow \infty$ \\
\end{tabular}
\end{ruledtabular}
\end{table}
%

Now investigate the solutions of Eq.(\ref{gamma-eq1}) in detail.
Differentiating Eq.(\ref{gamma-eq1}) with respect to $\ln x$, 
we get the following closed second order differential equation for $\gamma$, 
\bea
\gamma^{\prime\prime}-\frac{3-\alpha}{\alpha}\gamma^\prime
\left[\gamma-\frac{1}{3-\alpha}
\big(\gamma_a+\gamma_b-(\alpha-1)\gamma_1\big)\right]\nonumber\\
=\frac{\alpha-1}{\alpha^2}(\gamma-\gamma_a)(\gamma-\gamma_b)(\gamma-\gamma_1).
\label{gamma-eq2}
\eea
Using Eq.(\ref{gamma-eq2}),
we study the flow of solutions 
in the ($\gamma^\prime, \gamma$) phase space 
for the above three cases.
Note that the solution with $\gamma^\prime<0$ is unphysical 
because this solution is unstable.

\begin{figure}[h]
\includegraphics[width=8cm]{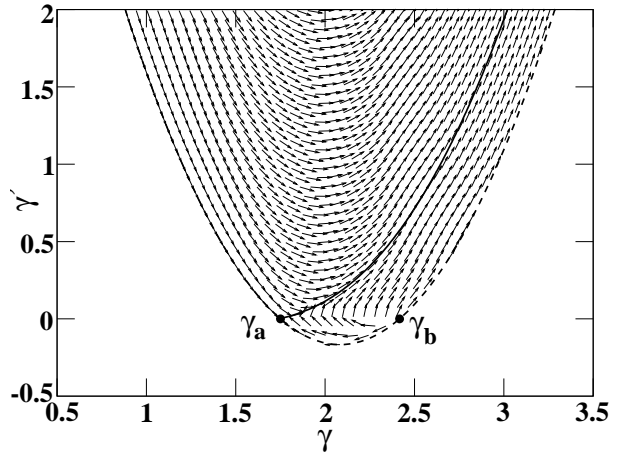}
\caption{\label{flow-A}
The flow diagram in the ($\gamma^\prime, \gamma$) phase space 
for case A.
The parameters are set to $(\beta,b)=(7/8,3/2)$.
A dashed line represents the boundary defined 
by Eq.(\ref{gamma-condition}).
Two filled circles denote the fixed points 
$\gamma_a$ (left) and $\gamma_b$ (right), respectively.
}
\end{figure}

In the case A, 
the flow of the solutions in Eq.(\ref{gamma-eq2}) is shown 
in Fig.\ref{flow-A}.
All solutions have an outer truncation beyond a finite radius
($\gamma\rightarrow \infty$).
The behavior of the solutions is classified into two families 
by the solution which starts from the fixed point $\gamma=\gamma_a$
and is represented by a solid line in Fig.\ref{flow-A}.
Solutions which exist in upper region of this solid line, are
that the density profile has an inner hole 
and an outer truncation.
These are unphysical solutions 
because the solutions with an inner hole are unstable.
On the other hand, 
solutions which exist in lower region of this solid line, are 
that the density profile behaves 
$\rho\propto r^{-\gamma_b}$ as $r\rightarrow 0$ and 
has an outer truncation.
This second family is also unphysical 
because this family has negative $\gamma^\prime$.
Finally, 
the solution which is represented by a sold line in Fig.\ref{flow-A} 
is realistic, 
because the density profile behaves 
$\rho\propto r^{-\gamma_a}$ as $r\rightarrow 0$ and 
has an outer truncation beyond a finite radius.

\begin{figure}[h]
\includegraphics[width=8cm]{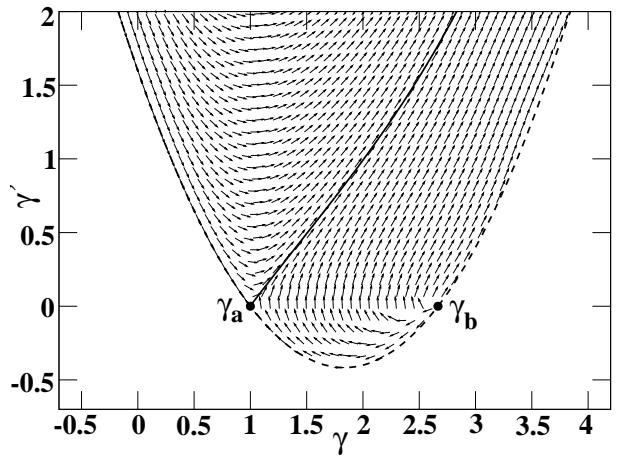}
\caption{\label{flow-C}
The same as Fig.\ref{flow-A}, 
but for case C.
The parameters are set to $(\beta,b)=(1/2,3/2)$.
}
\end{figure}

In the case C, 
the behavior of the $(\gamma^\prime, \gamma)$ phase space 
is shown in Fig.\ref{flow-C} and is same as one in the case A.
The realistic solutions are represented by a sold line where
the density profile behaves 
$\rho\propto r^{-\gamma_a}$ as $r\rightarrow 0$ and 
has an outer truncation.

\begin{figure}[h]
\includegraphics[width=8cm]{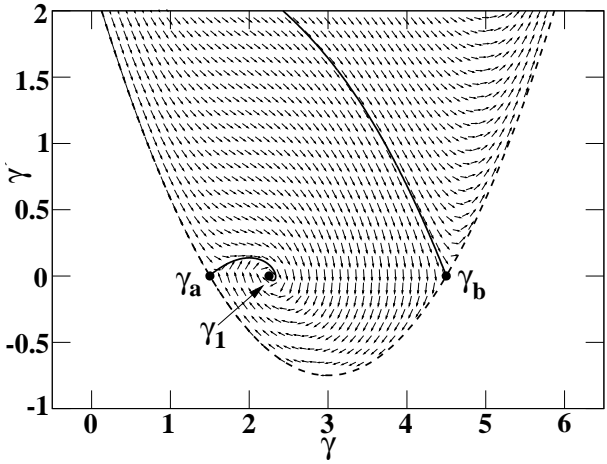}
\caption{\label{flow-B1}
The same as Fig.\ref{flow-A}, 
but for case B ($0<b<1$).
The parameters are set to $(\beta,b)=(3/4,3/4)$.
Three filled circles denote the fixed points 
$\gamma_a$ (left), $\gamma_1$(middle), and $\gamma_b$ (right), 
respectively.
}
\end{figure}
\begin{figure}[h]
\includegraphics[width=8cm]{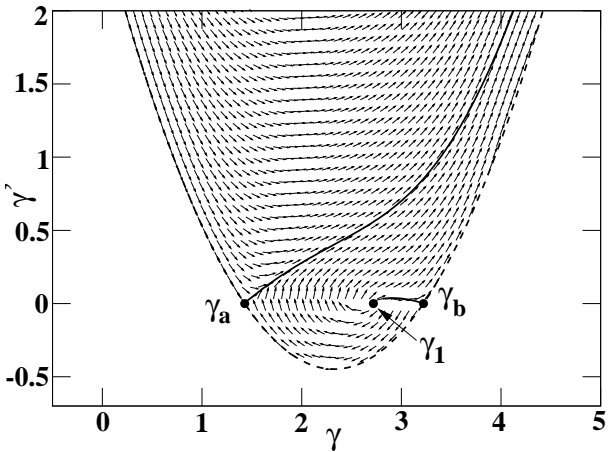}
\caption{\label{flow-B3}
The same as Fig.\ref{flow-B1}, 
but for case B ($1<b<2$).
The parameters set are to $(\beta,b)=(5/7,9/8)$.
}
\end{figure}
\begin{figure}[h]
\includegraphics[width=8cm]{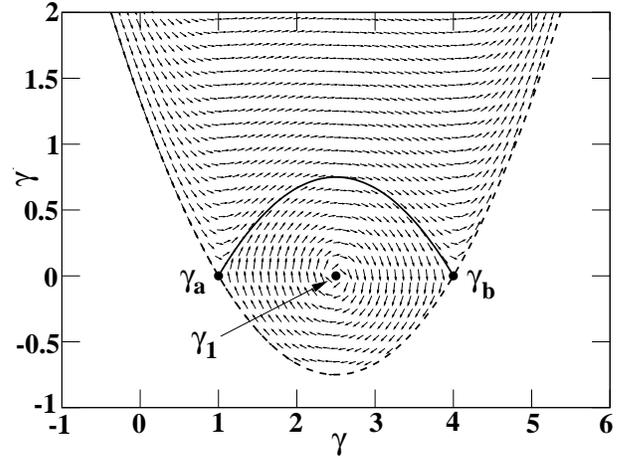}
\caption{\label{flow-B2}
The same as Fig.\ref{flow-B1}, 
but for case B ($b=1$).
The parameters are set to $(\beta,b)=(1/2,1)$.
The solid line connected between $\gamma_a$ and $\gamma_b$
denotes the critical solution.
}
\end{figure}

Figs.\ref{flow-B1}, \ref{flow-B3}, and \ref{flow-B2} show
the behavior of the $(\gamma^\prime, \gamma)$ phase space 
in the case B.
In this case, 
the behavior of the solutions in Eq.(\ref{gamma-eq2}) 
depends on whether $b$ is less than, greater than, 
or equal to a critical value $b_{\mbox{crit}}=1$, 
where $\gamma_1=(\gamma_a+\gamma_b)/2$ 
and the LV relation is satisfied.

For $b<b_{\mbox{crit}}$, 
the flow of the solutions is shown in Fig.\ref{flow-B1}.
In this case, 
all solutions except one are unphysical 
because the density profile has an inner hole.
The only exceptional solution is represented by a solid line 
which starts from $\gamma=\gamma_a$ and 
ends at $\gamma=\gamma_1$ in Fig.\ref{flow-B1}.
The density profile of this solution behaves as 
$\rho\propto r^{-\gamma_a}$ in the limit of $r\rightarrow 0$ and 
undergoes damped oscillation around $\rho\propto r^{-\gamma_1}$ 
in the limit of $r\rightarrow \infty$.

For $b>b_{\mbox{crit}}$, 
the flow of the solutions is shown in Fig.\ref{flow-B3}.
The behavior of the solutions in this case corresponds to 
a mirror image of one in the case $b<b_{\mbox{crit}}$.
The behavior of the solutions is classified into two families 
by the solution which starts from the fixed point $\gamma=\gamma_a$
and is represented by a solid line in Fig.\ref{flow-B3}.
Solutions which exist in upper region of this solid line, are unphysical 
because the solutions have an inner hole.
On the other hand, 
solutions which exist in lower region of this solid line, 
go around $\rho\propto r^{-\gamma_1}$ and 
have an outer truncation beyond a finite radius or 
end at $\gamma=\gamma_b$ ($\rho\propto r^{-\gamma_b}$).
This second family with positive $\gamma^\prime$ 
is physical.
Similarly, 
the solution which starts from $\gamma=\gamma_a$ 
and is represented by a sold line in Fig.\ref{flow-B3}, 
is also realistic, 
because the density profile behaves 
$\rho\propto r^{-\gamma_a}$ as $r\rightarrow 0$ and 
has an outer truncation beyond a finite radius.

Fig.\ref{flow-B2} shows the flow of the solution 
for $b=b_{\mbox{crit}}$.
In this case, 
there is a following first integral, 
\bea
K=&&\left[\gamma^\prime+\frac{2(1-\beta)}{5-4\beta}
    (\gamma-\gamma_a)(\gamma-\gamma_b)\right]^{2(1-\beta)}\nonumber\\
  &&\times\left[\gamma^\prime-\frac{1}{5-4\beta}
    (\gamma-\gamma_a)(\gamma-\gamma_b)\right].
\label{K}
\eea
Especially, 
$K=0$ leads to the Eq.(\ref{gamma-condition}) and 
\be
\gamma^\prime = -\frac{2(1-\beta)}{5-4\beta}
                (\gamma-\gamma_a)(\gamma-\gamma_b).
\label{critical-sol}
\ee
The above equation (\ref{critical-sol}) shows the critical solution, 
which starts from $\gamma=\gamma_a$ and 
ends at $\gamma=\gamma_b$ and is represented by a solid line 
in Fig.\ref{flow-B2}.
Solutions which exist in upper region of this critical solution, 
have an inner hole.
Others which exit in lower region of the critical solution, 
eternally oscillate around the fixed point $\gamma=\gamma_1$.
From these behaviors, 
all solutions are unphysical except the critical solution.
The only critical solution is physical and 
has a special characteristic that it shows power law behavior 
in the both limiting region 
$r\rightarrow 0$ ($\rho\propto r^{-\gamma_a}$) 
and $\infty$ ($\rho\propto r^{-\gamma_b}$).

It is easy to obtain an analytical form of the critical solution.
From Eq.(\ref{critical-sol}), 
we get 
\be
\gamma=\frac{\gamma_a+\gamma_b x^{2(1-\beta)}}{1+x^{2(1-\beta)}},
\label{gamma-cr}
\ee
where we choose an integral constant as $\gamma(1)=\gamma_1$.
Integrating again Eq.(\ref{gamma-cr}), 
we obtain
\bea
\rho &=&
\frac{1+s}{4\pi r_0^3}M_{tot}x^{2-s}(1+x^s)^{-(1+1/s)},
\label{rho-cr}\\
M &=&
M_{tot}(1+x^{-s})^{-(1+1/s)},\label{m-cr}\\
\phi &=&
\frac{GM_{tot}}{r_0}(1+x^s)^{-1/s},
\label{phi-cr}\\
\sigma^2 &=&
\frac{GM_{tot}}{2r_0}(1+x^s)^{-1/s},\label{sigma-cr}\\
\rho/\sigma^3 &=&
\frac{1+s}{\sqrt{2}\pi}\left(\frac{M_{tot}}{Gr_0}\right)^{3/2}\!\!\!
x^{2-s}(1+x^s)^{-(2-1/2s)},
\label{phase-cr}
\eea
where $s:=2(1-\beta)$, $M:=\int_0^x 4\pi u^2\rho du$, and 
$M_{tot}$ is a total mass $(4\pi r_0^3\rho_0)/(1+s)$.
This critical solution was found by \"U.-I.K. Veltmann \cite{Veltmann79} 
and N.W.Evans and J.An obtained by solving Jeans equation 
assuming the LV relation \cite{Evans05}.

The role of the LV ratio is similar to the scale invariant phase space density 
which is observed in the cosmological simulations 
based on the cold dark matter scenario.
Assuming the scale invariant phase space density 
($\sigma^3/\rho\propto r^{-a}$) and solving Jeans equation 
for spherically steady state system\cite{Dehnen05},
the critical value $a_{cr}=35/18$ exists and 
the flow diagram in the phase space around the critical value is 
similar to one in the case with the LV ratio.
The critical solution with the anisotropy parameter $\beta=7/8$ shows 
the scale invariant phase space density ($a=7/4\le a_{cr}$).

\section{\label{sec:comparison}Comparison with simulation}

\begin{figure}[h]
\begin{center}
\includegraphics[width=8cm]{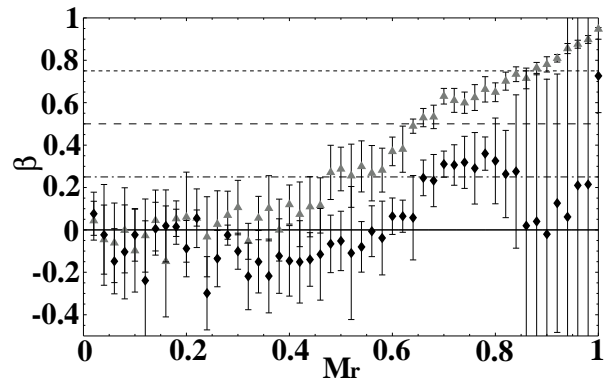}
\end{center}
\caption{The distribution of the anisotropy parameter $\beta(r)$
for the simulations of class I.
A plot with error-bar is the result of 
the numerical simulations SC with $(\bar{b_t},a)=(0,0)$ (gray triangle)
and $(\bar{b_t},a)=(0.5,0)$ (black diamond) in Fig.\ref{fig-cold}.
A solid horizontal line represents 
the critical solution (\ref{rho-cr})-(\ref{phase-cr}) with constant $\beta$. 
Each line represents the critical solution with $\beta=0$ (solid line), 
$0.25$ (dot-dashed line), $0.5$ (dashed line), 
and $0.75$ (dotted line), respectively.
}
\label{fbmcl1}
\end{figure}

\begin{figure}[h]
\begin{center}
\includegraphics[width=8cm]{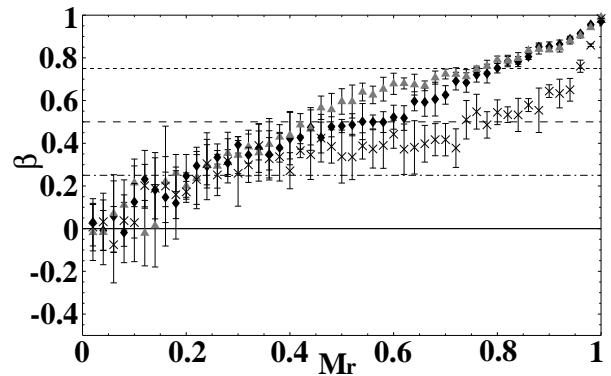}
\end{center}
\caption{Same as Fig.\ref{fbmcl1} but
for the simulations of class II.
A plot is the result of the numerical simulations SC 
with $(\bar{b_t},a)=(0,1)$ (black diamond),
$(0,2)$ (gray triangle), and CC (cross) in Fig.\ref{fig-cold}.
Each line represents the critical solution
with $\beta=0$ (solid line), $0.25$ (dot-dashed line), 
$0.5$ (dashed line), and $0.75$ (dotted line), respectively.
}
\label{fbmcl2}
\end{figure}

In the previous section, 
we showed that the LV relation is satisfied quite well 
for cold collapse simulations except for the case with 
a highly concentrated initial matter distribution. 
We can classify the N-body simulations into two main classes 
from the viewpoints of the behavior of anisotropy parameter $\beta$ 
for the bound state after a cold collapse. 
First is the ones with the initial homogeneous sphere 
where the particles are distributed homogeneously in the spherical region, 
which we call class I. 
In these initial conditions, 
large amount of particles reach the mass-center at the same time, 
which causes the high density central region with 
the isotropic velocity dispersion after the collapse. 
Actually we got the results that 
$\beta$ vanishes in the region within the central 30 percent cumulative mass 
regardless of the initial virial ratio in such cases  (Fig.\ref{fbmcl1}). 
On the other hand, 
the bound state starting from the other initial conditions show that 
$\beta$ is positive and monotonically increases 
against the cumulative mass (Fig.\ref{fbmcl2}). 
Hence we classify the simulations with these initial conditions 
into the second class and call it class II. 

\begin{figure}[h]
\begin{center}
\includegraphics[width=8cm]{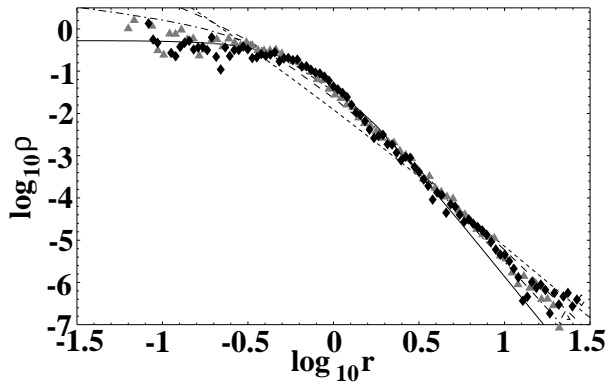}
\end{center}
\caption{A log-log plot for a density profile 
for class I simulations with the unit of $r_{h}=M_{tot}=G=1$, 
where $r_{h}$ is the half-mass radius of the bound system. 
A plot is the result of the numerical simulations SC 
with $(\bar{b_t},a)=(0,0)$ (gray triangle)
and $(\bar{b_t},a)=(0.5,0)$ (black diamond) in Fig.\ref{fig-cold}. 
Each line represents the critical solution 
with $\beta=0$ (solid line), $0.25$ (dot-dashed line), 
$0.5$ (dashed line), and $0.75$ (dotted line), respectively.
}
\label{frcl1}
\end{figure}

\begin{figure}[h]
\begin{center}
\includegraphics[width=8cm]{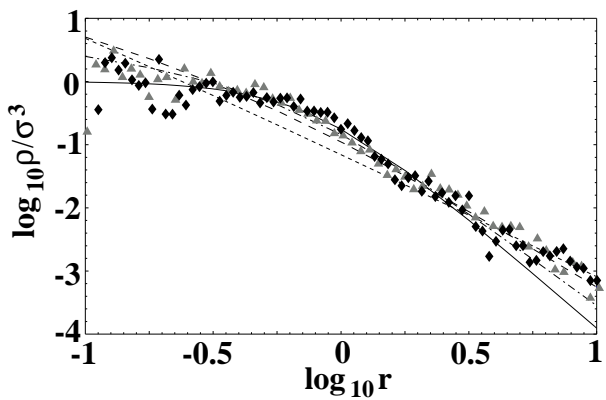}
\end{center}
\caption{Phase space density for class I simulations 
with the unit of $r_{h}=M_{tot}=G=1$.
A plot is the result of the numerical simulations SC 
with $(\bar{b_t},a)=(0,0)$ (gray triangle)
and $(\bar{b_t},a)=(0.5,0)$ (black diamond) in Fig.\ref{fig-cold}.
Each line represents the critical solution
with $\beta=0$ (solid line), $0.25$ (dot-dashed line), 
$0.5$ (dashed line), and $0.75$ (dotted line), respectively.}
\label{fpcl1}
\end{figure}
\begin{figure}[h]
\begin{center}
\includegraphics[width=8cm]{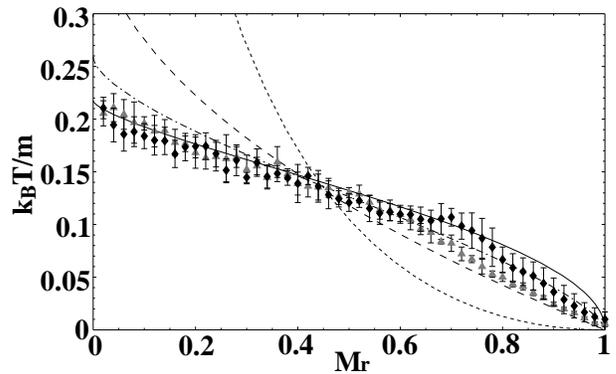}
\end{center}
\caption{Temperature-Mass relations for class I simulations 
with the unit of $r_{h}=M_{tot}=G=1$. 
A plot with error-bar is the result of 
the numerical simulations SC with $(\bar{b_t},a)=(0,0)$ (gray triangle)
and $(\bar{b_t},a)=(0.5,0)$ (black diamond) in Fig.\ref{fig-cold}.
Each line represents the critical solution
with $\beta=0$ (solid line), $0.25$ (dot-dashed line), 
$0.5$ (dashed line), and $0.75$ (dotted line), respectively.
}
\label{ftmcl1}
\end{figure}

\begin{figure}[h]
\begin{center}
\includegraphics[width=8cm]{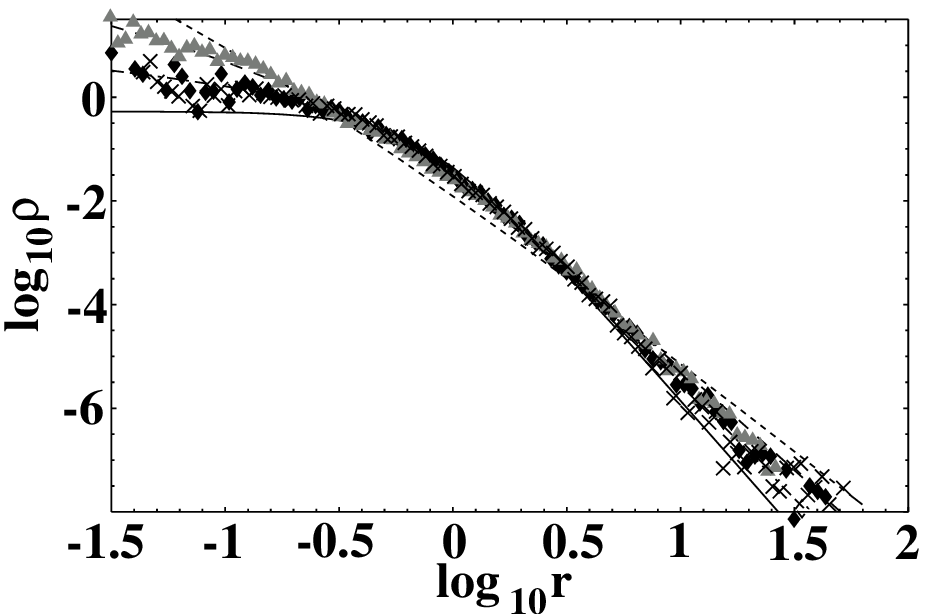}
\end{center}
\caption{Same as Fig.\ref{frcl1} but for class II simulations.
A plot is the result of the numerical simulations SC 
with $(\bar{b_t},a)=(0,1)$ (black diamond),
$(0,2)$ (gray triangle), and CC (cross) in Fig.\ref{fig-cold}.}
\label{frcl2}
\end{figure}

\begin{figure}[h]
\begin{center}
\includegraphics[width=8cm]{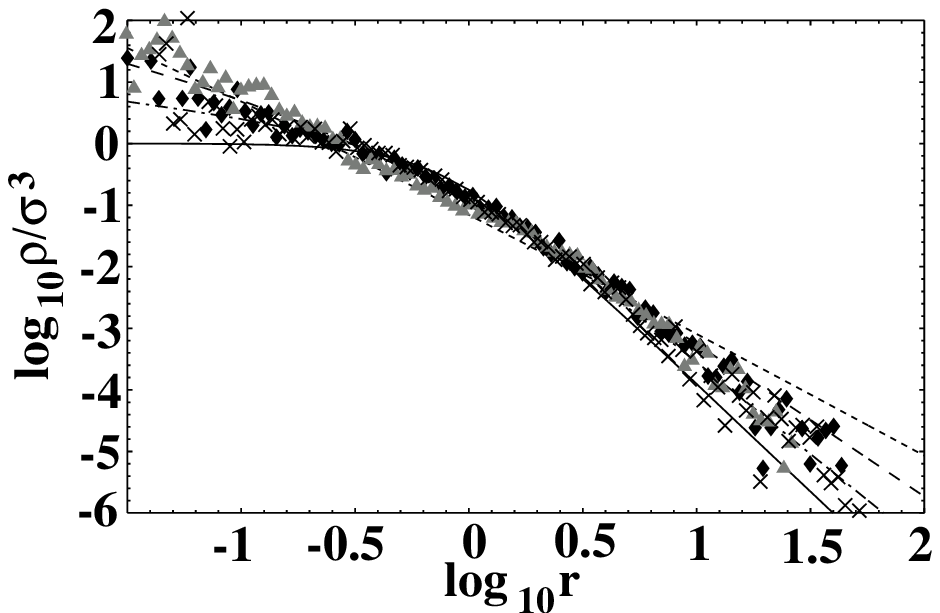}
\end{center}
\caption{Same as Fig.\ref{fpcl1} but for class II simulations.
A plot is the result of the numerical simulations SC 
with $(\bar{b_t},a)=(0,1)$ (black diamond),
$(0,2)$ (gray triangle), and CC (cross) in Fig.\ref{fig-cold}.}
\label{fpcl2}
\end{figure}

\begin{figure}[h]
\begin{center}
\includegraphics[width=8cm]{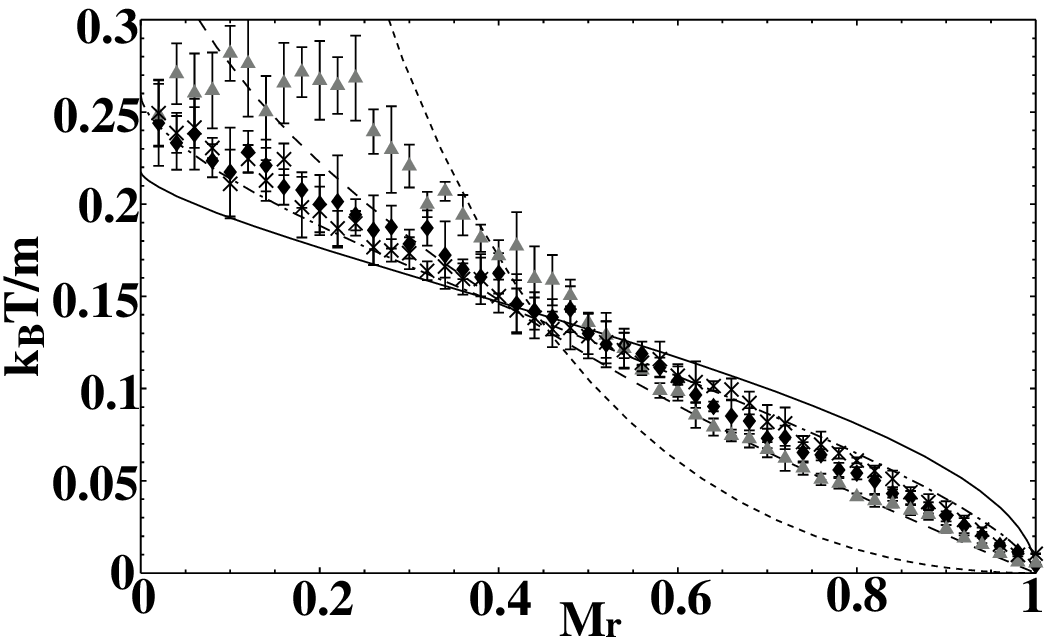}
\end{center}
\caption{Same as Fig.\ref{ftmcl1} but for class II simulations.
A plot with error-bar is the result of 
the numerical simulations SC with $(\bar{b_t},a)=(0,1)$ (black diamond),
$(0,2)$ (gray triangle), and CC (cross) in Fig.\ref{fig-cold}.}
\label{ftmcl2}
\end{figure}

Here we compare the results of these N-body simulations 
with those of the critical solutions in the previous section 
from the viewpoints of mass density, 
temperature-mass (TM) relation, and phase space density. 
In cosmological simulations, 
it is well known that the phase space density has 
a remarkable character, i.e., it has a single exponent 
with radius in the wide range of the viriallized region, 
despite the fact that the density has two different exponents \cite{Taylor01}. 
However, 
it is not obvious that it behaves in the same way 
for the simulations with different initial conditions. 
Hence it seems meaningful to examine the behavior of 
this quantity in our cold collapse simulations. 

First we compare the numerical results in the simulations of class I 
with the critical solutions with constant $\beta$. 
The densities fit well to the distribution 
of the Plummer's solution in the inner region 
as is shown in our previous paper (Fig.\ref{frcl1}). 
They, however, deviate from the Plummer's distribution 
in the outer region where the velocity anisotropy is developed. 
We observe that the density in this region fits well 
rather to that of the critical solution with positive constant $\beta$. 
Both the TM relation and the phase space density have 
the same characters, i.e., both of them fit well 
to those of the Plummer's solution in the inner region 
and to those of the positive constant $\beta$ critical solution 
in the outer region (Figs.\ref{fpcl1} and \ref{ftmcl1}). 
We can also derive the analytical solution 
which fits well to all of these physical quantities 
in the full region by connecting 
the inner Plummer's solution with the outer critical solution 
with positive constant $\beta$ (Appendix \ref{sec:cnnect} or reference \cite{Sota05}). 
The phase space density in these simulations 
becomes flat in the central part as well as the mass density. 
In the critical solution, 
the central part is described by the fixed point $\gamma=\gamma_a$, 
for which the exponent of the relative potential vanishes. 
Hence the exponent of the phase space density inevitably 
takes the same value as that of the mass-density at the center, 
as long as the phase space distribution follows 
the critical solution with constant $\beta$.

For the simulations of class II, 
the anisotropy parameter $\beta$ behaves quite differently. 
In those cases the velocity dispersion is anisotropic even 
in the inner region. 
Actually, 
the mass densities for those simulations are 
neither cuspy nor flat in the central region (Fig.\ref{frcl2}). 
Although the anisotropy parameter $\beta $ is 
not constant but is monotonically increasing outward, 
the density profiles fit well to that of the critical solutions 
with $\beta $=0.5 to 0.75 in full of the bound region. 
Both the phase space densities and TM relations 
for these simulations are also fitted to 
these critical solutions quite well, 
except for the case with the initial high density simulation with $(\bar{b_t},a)=(0,2)$. 

In this last case with $(\bar{b_t},a)=(0,2)$, 
the phase space density becomes cuspy against 
the behavior of the mass density (Fig.\ref{fpcl2}) and the temperature falls off 
at the center against the behavior of any critical solutions (Fig.\ref{ftmcl2}). 
These results are certainly correlated with the deviation 
from the LV relation at the central region. 
In this simulation, 
the kinetic energy is not sufficiently gained to attain the LV relation, 
which reflects both the temperature falling off 
and the steepness of the phase space density 
which is derived from dividing the density with the falling temperature.

\section{\label{sec:con}Conclusions and Discussions}

We focused on the local virial (LV) relation and 
investigated the role of the LV relation 
in a spherically symmetric steady state system.
Assuming the constant LV ratio and 
solving Jeans equation for spherically steady state system, 
we studied the full solution space of the problem 
under constant anisotropy parameter and 
obtained some meaningful solutions.
Especially, 
if the LV relation is satisfied, 
the solution of the power law behavior of 
the density profile in both the asymptotic regions 
$r\rightarrow 0$ and $\infty$ exists and 
is commonly observed in many numerical simulations.
In this sense, 
the LV relation plays a critical role in solutions of Jeans equation 
for spherically steady state system.

In the previous section, 
we compared our cold collapse simulations
with the critical solution 
with constant anisotropy parameter $\beta$. 
We got the results that the simulations agree with 
the critical solutions quite well except for 
the initially high density simulation with  $(\bar{b_t},a)=(0,2)$. 
In that simulation, 
the LV ratio $b$ becomes smaller than one around the center, 
where the temperature falls off and the phase-space density 
becomes steeper than the mass density. 

These results remind us of the cosmological simulations 
for which both mass density and the phase space density 
have the universal characters. 
In the cosmological simulation, 
the phase space density behaves as $\rho/\sigma^3 \sim r^{-\alpha}$, 
where $\alpha \sim 1.87$ \cite{Dehnen05,Taylor01}, 
which is larger than the exponent of the mass density in the central part
and the temperature falls off in the central region. 
Hence we speculate that both of the initial high density simulations 
and cosmological simulations have the common characters 
for the phase space distribution. 
This may be because in the initially high density simulation, 
the matters in the central part collapse to make a core 
in the early stage and the matters 
in the outer part gradually falls into the core in the later stage, 
which is similar to the case with the secondary in-fall 
in the cosmological simulation.

However, 
we have to be careful to conclude that 
these results mean the inapplicability of the LV relation 
for these cases, because we compared these cases only 
under the condition $\beta$ is constant. 
It is not obvious that the same kind of asymptotic behavior of 
the critical solutions obtained under the LV relation 
are sustained for any function form of $\beta (r)$. 
We will discuss these points in our coming paper.

\appendix
\section{A solution for $\alpha=1$ case}
\label{sec:alpha1}
\subsection{$\beta=1$ case}
In the case for $(\beta,b)=(1,1)$, 
the equation (\ref{gamma-eq1}) becomes 
\be
\gamma^{\prime}=(\gamma-\gamma_{+})(\gamma-\gamma_{-})=\kappa,
\label{gamma1-alpha1}
\ee
where 
\be
\gamma_{+}=\frac{5+\sqrt{1-4\kappa}}{2}, \quad\quad
\gamma_{-}=\frac{5-\sqrt{1-4\kappa}}{2}.
\ee
Integrating the Eq.(\ref{gamma1-alpha1}), 
we obtain
\be
\gamma=\frac{\gamma_{+}+\gamma_{-}x^{\gamma_{+}-\gamma_{-}}}
{1+x^{\gamma_{+}-\gamma_{-}}},
\label{gamma-alpha1-sol}
\ee
where we choose the integral constant 
as satisfying the condition $\gamma(1)=(\gamma_{+}+\gamma_{-})/2$.
At large $x$, 
$\gamma$ approaches $\gamma_{-}$ and 
the total mass is infinite because $\gamma_{-}\le 5/2$.
From Eqs.(\ref{Phi-rho}), (\ref{Jeans}), 
and (\ref{gamma-alpha1-sol}), 
we have
\bea
&&\rho\propto
x^{-\gamma_{+}}\left(1+x^{\gamma_{+}-\gamma_{-}}\right),\\
&&M\propto
4\pi x^{3-\gamma_{+}}\left(\frac{1}{3-\gamma_{+}}
+\frac{x^{\gamma_{+}-\gamma_{-}}}{3-\gamma_{-}}\right),\\
&&\phi\propto
x^{2-\gamma_{+}}\left(1+x^{\gamma_{+}-\gamma_{-}}\right),\\
&&\sigma^2\propto
\frac{1}{2}x^{2-\gamma_{+}}\left(1+x^{\gamma_{+}-\gamma_{-}}\right),\\
&&\rho/\sigma^3\propto
2^{3/2}x^{\gamma_{+}/2-3}\left(1+x^{\gamma_{+}-\gamma_{-}}\right)^{-1/2},
\label{alpha1-sol}
\eea
where $M:=\int_0^x 4\pi u^2\rho du$.

\subsection{$\beta\neq 1$ case}

\begin{figure}[h]
\includegraphics[width=8cm]{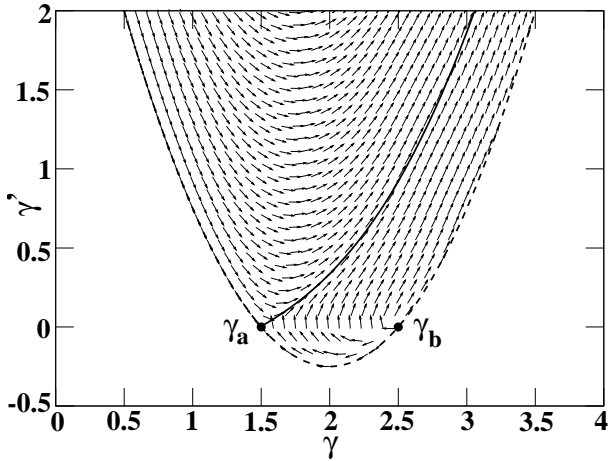}
\caption{\label{flow-alpha1}
The same as Fig.\ref{flow-A}, 
but for case $\alpha=1 (\beta\neq 1)$.
The parameters are set to $(\beta,b)=(3/4,3/2)$.
}
\end{figure}

In this case, 
the equation (\ref{gamma-eq1}) becomes 
\be
\gamma^{\prime}-(\gamma-\gamma_a)(\gamma-\gamma_b)=
\kappa x^{2(1-\beta)},
\label{gamma-alpha1-eq1}
\ee
where
\be
\gamma_a=2\beta,\quad\quad
\gamma_b=2\beta+1.
\ee
The behavior of the solutions in Eq.(\ref{gamma-alpha1-eq1}) 
shows in Fig.\ref{flow-alpha1}.
In the limit $r\rightarrow 0$, 
$\gamma$ approaches $\gamma_a$ or $\gamma_b$ or $-\infty$ (inner hole).
On the other hand, 
in the limit  $r\rightarrow \infty$, 
all solutions have an outer truncation at $r=\infty$.
Physically meaningful solution is represented by a solid line in
Fig.\ref{flow-alpha1} where
$\gamma$ approaches $\gamma_a$ at $r\rightarrow 0$.

\section{the connection of relative potentials}
\label{sec:cnnect}

Here we connect two critical solutions 
with different values of constant $\beta$, 
which leads to the critical solution 
with a step function form of $\beta \left( r \right)$.
Here we use the unit $G=r_c =\phi _c =1$, 
where $r_c$ is the radius of the connected position and 
$\phi_c$ is the value of the relative potential at $r=r_c$.
In this unit¡¤
the relative potentials $\phi_1$ in the inner region and 
$\phi_2$ in the outer region are described as
\bea
\phi_1 &=&\left( \frac{1+c_1^{s_1}}{r^{s_1}+c_1^{s_1}}\right)^{1/s_1},\\
\phi_2 &=&\left( \frac{1+c_2^{s_2}}{r^{s_2}+c_2^{s_2}} \right)^{1/s_2},
\eea
respectively. 
Each potential includes a free parameter $c_1$ or $c_2$
other than the parameter $s_1$ and $s_2$ related to the
anisotropy of each region.
From the continuous condition of the first derivative of 
$\phi_1$ and $\phi_2 $ against $r$ at $r=1$,
$c_2 $ is described as
\bea
c_2 =c_1^{s_1/s_2}.
\eea
Hence connected solution depends on 
the one-parameter $c=c_1 $ for given $s_1 $ and $s_2 $.

This parameter $c$ determines the mass fraction of 
the inner solution $\phi_1$ against total mass as
\bea
\frac{M_1 }{M_{tot} }=\left( {1+c^{s_1}} \right)^{-1-1/s_2}.
\eea
Hence we can adjust the mass fraction of the inner region 
by changing the parameter $c$.


\begin{acknowledgments}
The authors would like to thank Professor Kei-ichi Maeda for the extensive discussions.
All numerical simulations were carried out on GRAPE system at ADAC (the
Astronomical Data Analysis Center) of the National Astronomical Observatory,
Japan.
\end{acknowledgments}



\end{document}